\documentstyle[multicol,aps,psfig]{revtex}
\renewcommand{\narrowtext}{\begin{multicols}{2}
\global\columnwidth20.5pc\noindent}
\renewcommand{\widetext}{\end{multicols}
\global\columnwidth42.5pc}
\def\vS{\mbox{\boldmath{$S$}}}
\def\vs{\mbox{\boldmath{$s$}}}

\begin{document}
\draft
\preprint{April 17, 2002}
\title{Quantum magnetization plateaux
of an anisotropic ferrimagnetic spin chain
       }
\author{T$\hat{\mbox o}$ru Sakai}
\address
{Tokyo Metropolitan Institute of Technology,
6-6 Asahigaoka, Hino, Tokyo 191-0065, Japan \\
Department of Physics, Tohoku University, Aramaki, 
Aoba-ku, Sendai 980-8578, Japan}
\author{Kiyomi Okamoto}
\address
{Department of Physics, Tokyo Institute of Technology,
 Meguro-ku, Tokyo 152-8551, Japan}
%\date{Received \hspace{6cm}}
\date{Received April 17, 2002}
\maketitle
\begin{abstract}
The magnetization curve of the $(S,s)=(1,1/2)$ ferrimagnetic alternating
spin chain with the single-ion anisotropy $D$ is investigated
with the numerical exact diagonalization of finite clusters and
size-scaling analyses.
The system has a plateau at 1/3 of the saturation moment,
which corresponds to the spontaneous magnetization for $D=0$.
Varying $D$ in the 1/3-magnetized ground state under
the external field along the axis of $D$,
a quantum phase transition is revealed to occur
at the critical value $D/J=1.114 \pm 0.001$ where the
plateau vanishes.
Except for the critical point,
the plateau is always opening, but the mechanism is
different between $D<D_c$ and $D>D_c$.
%The two plateau phases correspond to the Haldane and large-$D$
%phases in the $S=1$ antiferromagnetic chain.
The change of mechanisms is an evidence to clarify that
the plateau originates from the quantization of magnetization.

\end{abstract}
\pacs{PACS numbers: 75.10.Jm, 75.40.Mg, 75.50.Gg, 75.40.Cx}
\narrowtext

\section{Introduction}

The quantization of the magnetization is one of interesting phenomena
in low-dimensional magnets.
It is detected as a plateau in the magnetization curve.
Such plateaux were actually observed in high-field measurements
of several materials: the $S=1$ bond-alternating chain
[Ni$_2$(Medpt)$_2$($\mu$-ox)($\mu$-N$_3$)]ClO$_4\cdot$0.5H$_2$O
(Medpt=methyl-bis(3-aminopropyl)amine)\cite{hagiwara1,hagiwara2,tonegawa},
the organic $S=1$ ladder
3,3',5,5'-tetrakis($N$-$tert$-butylaminoxyl)biphenyl, abbreviated
BIP-TENO\cite{katoh,goto,okamoto1,okazaki,okamoto2},
the Shastry-Sutherland system SrCu$_2$(BO$_3$)$_2$
\cite{kageyama,miyahara}
and the zigzag double chain NH$_4$CuCl$_3$\cite{shiramura,kolezhuk} etc.
In addition some theoretical and/or numerical analyses predicted
that a magnetization plateau appears in various other systems;
the polymerized chains
\cite{hida,okamoto3,tonegawa2,totsuka1,totsuka2,cabra1,honecker1,%
oka-kita1,oka-kita2,oka-kita3,tonegawa3},
the $S=3/2$ chain\cite{oshikawa,sakai1,kitazawa1,kitazawa12},
the frustrated spin ladder\cite{mila,tandon,okazaki1,okazaki2,honecker2},
several generalized spin ladders
\cite{cabra2,cabra3,cabra4,cabra5,sakai2,nakasu},
distorted diamond type spin chain\cite{tonegawa4,tonegawa5,honecker*}
and some layered sytems\cite{honecker3,honecker4}.
Using the Lieb-Schultz-Mattis theorem\cite{lieb},
a general necessary condition for the presence of the plateau
was derived\cite{oshikawa} as
\begin{equation}
   \widetilde{S}-m=\mbox{integer}\,,
   \label{oya}
\end{equation}
where $\widetilde{S}$ and $m$ are the sum of
spins over all sites and the magnetization in the unit period, respectively.

The ferrimagnetic mixed spin chains have lately attracted a lot of
interest among quantum spin systems.
Recent synthesizing techniques have produced a lot of such materials,
for example, the bimetallic chain MM$'$(pbaOH)(H$_2$O)$_3\cdot$nH$_2$O
\cite{kahn}
and the organic one \{Mn(hfac)$_2$\}$_3$(3R)$_2$\cite{markosyan} etc.
Thus many experimental investigations have been done on such ferrimagnets,
as well as theoretical ones.
\cite{drillon,alcaraz,pati,brehmer,niggemann,yamamoto1,ono,kuramoto,%
yamamoto2,ivanov,maisinger,yamamoto3}
In the systems two different spins $S$ and $s$ $(S>s)$
are arranged alternately
in a line and coupled by the nearest-neighbor antiferromagnetic
exchange interaction.
They has the spontaneous magnetization $m=S-s$ in the ground state.
Since the lowest excitation increasing $m$ has an energy gap,
the magnetization curve has a plateau at $m=S-s$\cite{kuramoto}.
Indeed the plateau satisfies the condition of the quantization of the
magnetization (\ref{oya}).
The plateau, however, is also realized in the classical limit where
$S$ and $s$ are infinite with the ratio $S/s$ fixed.
Thus it is difficult to identify the plateau at $m=S-s$ as a
result from the quantization of the magnetization,
unlike any higher plateaux at $m=S-s+1, S-s+2, \cdots, S+s-1$
which should not appear in the classical Heisenberg spin systems
\cite{ys2,sy2}.
The previous works\cite{sy1,ys1} to investigate the anisotropy
in the exchange
interaction revealed that the quantum effect stabilizes the
plateau at $m=S-s$ against the $XY$-like anisotropy.
However, it is no more than a quantitative difference between the quantum
and classical systems.
In this paper,
to show a more definite evidence to clarify that the plateau
is a result from the quantization,
we investigate the single-ion anisotropy $D$ effect.
It is more realistic than the interaction anisotropy
even from an experimental viewpoint.
For example, the recently synthesized $(S,s)=(1,1/2)$ chain
NiCu(pba)(D$_2$O)$_3$$\cdot$2D$_2$O\cite{hagiwara3} would possibly have
the anisotropy $D$ on the Ni ion ($S=1$).
Thus we study on the (1,1/2) system with $D$ for $S=1$,
using the numerical diagonalization and the level spectroscopy method
\cite{oka-nomu,nomura1,nomura2,okamoto4}
under the twisted boundary condition\cite{kitazawa2,nomu-kita}.
It will reveal that the system has a quantum phase transition with
respect to $D$ at $m=1/2$ due to the change of plateau formation
mechanisms, which clarifies a full quantum nature of the plateau.

\section{Classical System}

\begin{figure}
\mbox{\psfig{figure=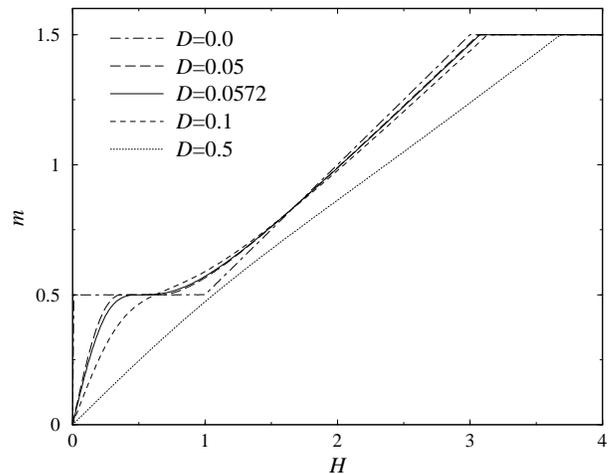,width=80mm,angle=-90}}
\caption{
Magnetization curves of the classical $(S,s)=(1,1/2)$ chain for
various $D$. The plateau appears at $m=1/2$ only for
$D<D_c=0.0572$.
}
\label{fig1}
\end{figure}

The ferrimagnetic $(S,s)=(1,1/2)$ mixed spin chain with the single-ion
anisotropy under an external magnetic field $H$ is described by the
Hamiltonian
\begin{eqnarray}
   {\cal H}
   &=& {\cal H}_0 + {\cal H}_{\rm ex}
      \label{ham} \\
   {\cal H}_0
   &=&\sum_{j=1}^N
     \left\{ \vS_{j}\cdot\vs_{j} + \vs_{j}\cdot\vS_{j+1}
          +D\left(S_j^z\right)^2 \right\}
   \label{ham0}  \\
   {\cal H}_{\rm ex}
   &=& -H \sum_{j=1}^N (S_j^z+s_j^z)
   \label{ham-ex}
\end{eqnarray}
Throughout this paper, we consider only the case when
the external field is along the symmetry axis of $D$.
We note that $D(s_j^z)^2$ gives a constant $D/4$.
In order to clarify the quantum nature of the plateau later,
we show the properties of the classical spin system
with the same Hamiltonian where $\boldmath S$ and $\boldmath s$
are the classical vectors with the amplitudes 1 and 1/2,
respectively.
The magnetization process at $T=0$ can be solved with the
standard variation of the Hamiltonian with respect to
the two vectors $\boldmath S$ and $\boldmath s$.
In the isotropic case of the classical system
the ground state magnetization curve begins at the spontaneous
magnetization $m=S-s=1/2$ and has a plateau there, shown as a
dot-dashed line in Fig. \ref{fig1}.
The complete N\'eel order along $H$ ($S_j^z=1, s_j^z=-1/2$) is
realized in the plateau state, while the canted N\'eel order
(in the $xy$ plane) occurs for $1/2<m<3/2$.
Thus the plateau should originate from the classical N\'eel order.
Since the N\'eel order is oriented in the $xy$ plane for $D>0$,
the magnetization curve starts from $m=0$.
The plateau at $m=1/2$ still appears for small positive $D$,
as a long dashed line ($D=0.05$) in Fig. \ref{fig1},
because the N\'eel order along $H$ can be realized there.
With increasing $D$, however, the plateau disappears at the
critical value $D_c=0.0572$ and it does not appear any more
for $D>D_c$, shown in Fig. \ref{fig1}.
The breakdown of the plateau due to the easy plane anisotropy $D>0$
is qualitatively the same as the case of the $XY$-like coupling
anisotropy\cite{sy1,ys1}.
In the quantum system discussed in the following sections, however,
the plateau will be revealed to appear again for larger $D$ due to
the quantum effect, in contrast to the coupling anisotropy.

\section{Quantum Mechanisms of Plateau}

In the quantum spin system described by the Hamiltonian (\ref{ham})
it would be efficient to introduce the composite spin picture,
where $S=1$ is considered as the triplet state of two 1/2 spins
\cite{aklt}.
Using this picture we clarify the two different mechanisms of the
plateau at $m=1/2$; the Haldane and large-$D$ mechanisms for
small and large $D$, respectively.
The names of these two mechanisms
originate from the Haldane and large-$D$ phases
of the $S=1$ antiferromagnetic chain
where the critical point was revealed to be $D \sim 1$
\cite{haldane,sakai3,golinelli}.
Similar mechanisms were also proposed for the 1/3 plateau of
the $S=3/2$ chain\cite{kitazawa1,kitazawa12}.

\subsection{Haldane Plateau}

\begin{figure}
\mbox{\psfig{figure=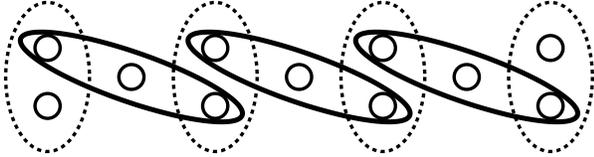,width=80mm,angle=0}}
\caption{
Schematic picture of the Haldane mechanism of the plateau at $m=1/2$.
Each circle represents a 1/2 spin.
A solid ellipse is a trimer and a dotted ellipse means
symmetrized two spins.
}
\label{fig2}
\end{figure}

For smaller $D$ each $S=1$ site can be in any state of $S^z=$-1, 0 and +1.
Thus the two 1/2 spins representing $S=1$ have only to be symmetrized.
In the plateau state at $m=1/2$ (1/3 of the saturation magnetization),
the antiferromagnetic trimer state,
which is schematically shown in Fig. \ref{fig2}, is expected to
be realized.
In Fig. \ref{fig2} a solid ellipse represents the trimer
\begin{eqnarray}
|\Uparrow \rangle = {1\over {\sqrt{6}}}(|\uparrow \uparrow \downarrow \rangle
-2|\uparrow \downarrow \uparrow \rangle
+ |\downarrow \uparrow \uparrow \rangle ),
\label{trimer}
\end{eqnarray}
and a dotted one corresponds to an $S=1$ site where the two 1/2 spins
should be symmetrized.
When the plateau is based on the effective mechanism in Fig. \ref{fig2},
we call it the Haldane plateau.
In the ideal state where this picture is exactly realized,
the expectation values of the $z$-component are ${1\over 3}, -{1\over 6}$
and ${1\over 3}$ for the three spins in each trimer, respectively.
Thus the original system is expected to have
$\langle S^z \rangle ={2\over 3}$ and $\langle s^z \rangle =-{1\over 6}$
in the ideal Haldane state.
It suggests that the Haldane state has the N\'eel order along $H$,
but the amplitude is smaller than the classical system
($\langle S^z \rangle =1$ and $\langle s^z \rangle =-{1\over 2}$)
because of the quantum spin reduction.
We note that the $D=-\infty$ case is not the ideal Haldane case.
In this case, classical values
$\langle S^z \rangle =1$ and  $\langle s^z \rangle =-1/2$ will be realized.

\subsection{Large-$D$ Plateau}

\begin{figure}
\mbox{\psfig{figure=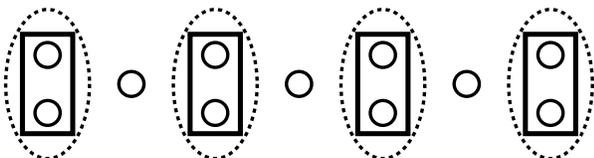,width=80mm,angle=0}}
\caption{
Schematic picture of the large-$D$ mechanism of the plateau at $m=1/2$.
Each circle represents a 1/2 spin.
A solid rectangle is
one of the triplet (6).
%${1\over {\sqrt{2}}} ( | \uparrow \downarrow \rangle
%+ | \downarrow \uparrow \rangle )$.
}
\label{fig3}
\end{figure}

Each $S=1$ site tends to have $S^z=0$ for larger positive $D$.
In the composite spin picture $S^z=0$ corresponds to one of the triplet
states
\begin{eqnarray}
{1\over {\sqrt{2}}}(|\uparrow \downarrow \rangle + |\downarrow \uparrow
\rangle )
\label{triplet}
\end{eqnarray}
Thus the large-$D$ mechanism of the 1/3 plateau ($m=1/2$) is presented
schematically in Fig. \ref{fig3}, where
a rectangle represents the triplet state (\ref{triplet}) at $S=1$ site
and each $s=1/2$ site has $s^z=1/2$.
In the ideal large-$D$ state the expectation values should be
$\langle S^z \rangle =0$ and $\langle s^z \rangle =+{1/ 2}$.
Obviously no N\'eel order is realized along $z$-axis
in the large-$D$ phase.
The large-$D$ plateau results from the quantization of $S=1$.
Thus it is never realized in the classical spin system.

\section{Quantum Critical Point}

\begin{figure}
\mbox{\psfig{figure=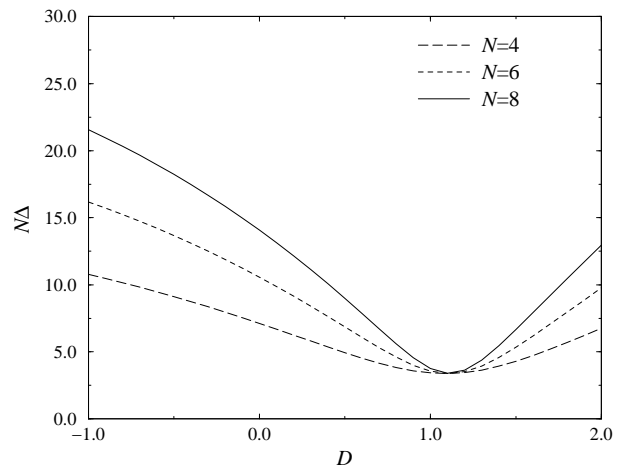,width=80mm,angle=-90}}
\caption{
Scaled gap $N\Delta$ plotted versus $D$.
It suggests that the critical point is around $D=1.1$.
}
\label{fig4}
\end{figure}

The existence of the two different mechanisms of the 1/3 plateau
suggests that there is a quantum phase transition between them
with respect to the parameter $D$ in the ground state at $m=1/2$.
A useful order parameter to investigate the phase transition
is the spin excitation gap at $m=1/2$
\begin{eqnarray}
\Delta \equiv E(M+1)+E(M-1)-2E(M).
\label{plateau}
\end{eqnarray}
E(M) is the lowest energy level in the subspace where
the eigenvalue of $\sum _j (S^z_j+s^z_j)$ is $M$ and
$m=1/2$ corresponds to $M=N/2$ for the $N$-unit system.
$\Delta$ is also the length of the plateau.
Based on the analogy with the $S=1$ chain,
it is expected that
the critical point $D_c$ between the Haldane and large-$D$ phases
is a Gaussian fixed point.
Thus the gap would vanish just at $D_c$ and open in both phases.
The scaled gap $N\Delta$ calculated for several finite systems under
the periodic boundary condition using the numerical diagonalization
is plotted versus $D$ in Fig. \ref{fig4}
It indicates that there exists a critical point at $D\sim 1.1$ where
the scaled gap $N\Delta$ is independent of $N$, namely the system is
gapless ($\Delta \sim 1/N$).
It also suggests that the gap is opening for both sides of $D_c$,
as expected.
%
% append 1
%
In general, the fixed point of the phenomenological renormalization
equation\cite{nightingale}
\begin{eqnarray}
N \Delta _N(D)=(N+2) \Delta _{N+2} (D')
\label{prg}
\end{eqnarray}
depends on the system size $N$ for small clusters.
However, the behavior of the scaled gap in Fig. \ref{fig4} suggests that
the fixed point is almost independent of $N$, that is
$D_{c, 4,6} \sim D_{c, 6,8}$.
It implies that the higher-order size correction ($o(1/N)$) of the gap
$\Delta$ is negligible even for $N=$4, 6, and 8 in the present system.
Thus some systematic size-scaling analyses would lead to a
precise estimation of the critical point in the thermodynamic limit,
even based only on such small cluster calculations.

\begin{figure}
\mbox{\psfig{figure=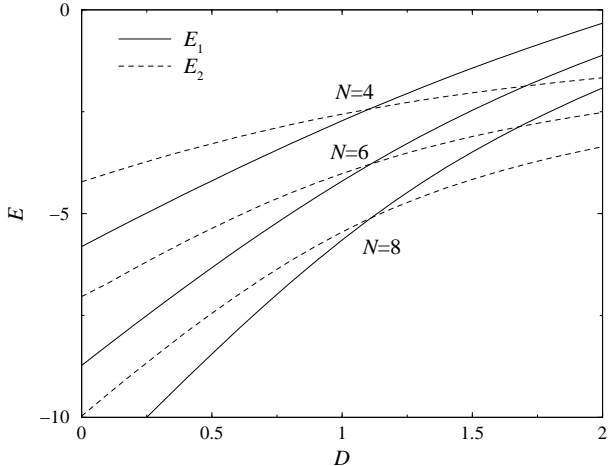,width=80mm,angle=-90}}
\caption{
Lowest two energy levels under the twisted boundary condition.
The crossing point is the size-dependent critical point.
}
\label{fig5}
\end{figure}

\begin{figure}
\mbox{\psfig{figure=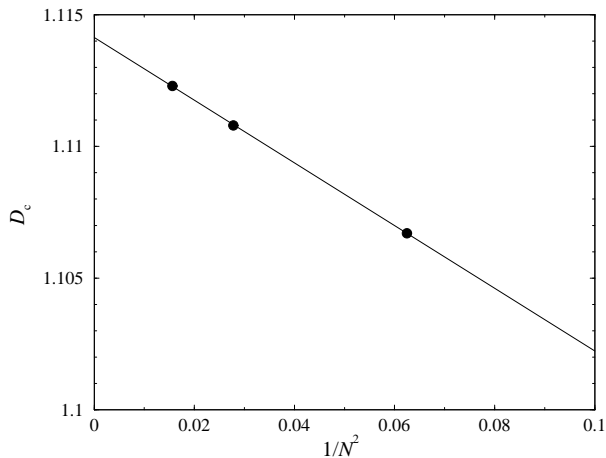,width=80mm,angle=-90}}
\caption{
Estimation of the critical point $D_c$ in the thermodynamic limit
based on fitting a line $\sim 1/N^2$.
}
\label{fig6}
\end{figure}

One of the most precise methods to determine the Gaussian fixed point
of the one-dimensional quantum systems is the level spectroscopy
\cite{oka-nomu,nomura1,nomura2,okamoto4}
under the twisted boundary condition\cite{kitazawa2,nomu-kita}.
The twisted boundary condition means that the sign of the coupling
constant for the $XY$ component is switched in the exchange
interaction at the boundary.
According to the method,
the critical point is determined as a crossing point
of the lowest two energy levels under twisted boundary conditions.
The two levels $E_1$ and $E_2$ are plotted versus $D$
for the finite system with $N=$4, 6 and 8 in Fig. \ref{fig5}.
The size dependence of the crossing point is so small that
a precise $D_c$ is expected to be obtained.
%
% append 2
%
Since the effect of some irrelevant fields\cite{nomura1}
should yield the
size correction proportional to $1/N^2$,
we extrapolate the size-dependent $D_c$ to the thermodynamic
limit using the plot versus $1/N^2$ in Fig. \ref{fig6}.
The result is $D_c=1.114 \pm 0.001$.
%
% append 3
%
The good agreement of the size correction with $1/N^2$
in Fig. \ref{fig6} justifies
the accuracy of the present estimation.
In fact, the plateau-nonplateau phase boundary of some spin ladder systems
determined by the level spectroscopy even with small cluster calculations
well agreed with the exact result in some ideal limits.
\cite{okamoto1,okazaki2,nakasu}
The above scaled gap analysis also supports the existence of the
quantum critical point in the present system.

\begin{figure}
\mbox{\psfig{figure=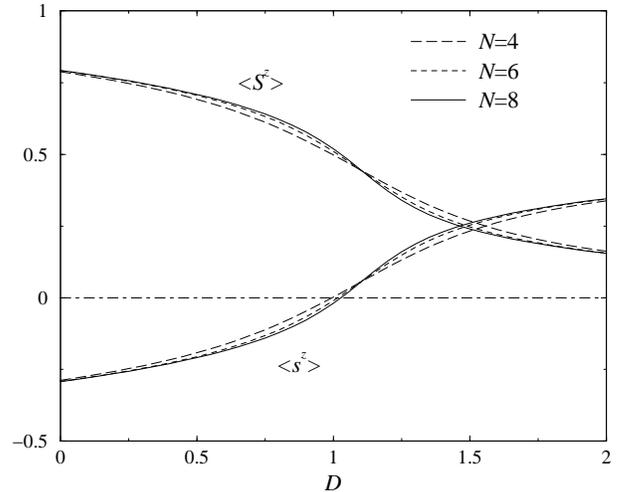,width=80mm,angle=-90}}
\caption{
Sublattice magnetizations $\langle S^z \rangle$ and
$\langle s^z \rangle$ plotted versus $D$.
It suggests that the sign of
$\langle s^z \rangle$ is changed at the critical point.
}
\label{fig7}
\end{figure}

In order to confirm the realization of the two mechanisms of the
plateau discussed in the previous section,
the expectation values of the $z$ component
$\langle S^z \rangle $ and $\langle s^z \rangle $ for finite
systems are plotted versus $D$ in Fig. \ref{fig7}.
It indicates that
the sign of $\langle s^z \rangle $ switched around the critical
point.
It means that the N\'eel order along $z$-axis exists only
in the Haldane phase.
Thus it is also consistent with the schematic pictures of the
two plateau mechanisms.

\section{Magnetization Curves}

\begin{figure}
\mbox{\psfig{figure=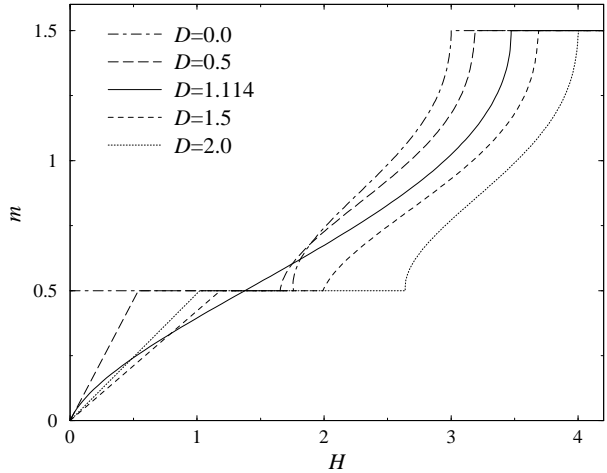,width=80mm,angle=-90}}
\caption{
Magnetization curves of the quantum system for several values of $D$.
The plateau at $m=1/2$ is caused by the Haldane mechanism for
$D=0$ and 0.5, while the large-$D$ one for $D=1.5$ and 2.0.
The curve has no plateau just at the critical value $D=1.114$.
}
\label{fig8}
\end{figure}

Finally, we present the ground state magnetization curve of the quantum
systems described by the Hamiltonian (\ref{ham}) for several values of $D$,
using some size scaling techniques\cite{sakai4}
applied for the numerical energy levels
of finite systems up to $N=10$.
The conformal field theory in one-dimensional quantum systems
\cite{cardy,blote,affleck} applied for the present model
predicted that the in gapless phases the size dependence of the energy gap
have the asymptotic forms
\begin{eqnarray}
E(M+1)-E(M) \sim H(m) + \pi v_s \eta {1\over N}, \\
E(M)-E(M-1) \sim H(m) - \pi v_s \eta {1\over N},
\label{gapless}
\end{eqnarray}
where $N$ and $M$ vary with $m=M/N$ fixed.
Thus the forms are useful to estimate the magnetic field $H$
for several values of $m$ which can be obtained from all the
combinations of $N$ and $M$ for available finite systems.
The method works except for the plateau.
On the other hand,
at the massive point like plateaux the Shanks transformation
\cite{shanks}, which is a technique to accelerate the convergence of
sequences, is useful.
The general form of the transformation for a sequence $a_n$ is
\begin{eqnarray}
a'_n ={{a_{n-1}a_{n+1}-a_n^2}\over {a_{n-1}+a_{n+1}-2a_n}}.
\label{shanks}
\end{eqnarray}
We show the result of the magnetization curve for several values of $D$
in Fig. \ref{fig8}, where only the polynomial curves suitably fitted to the
obtained points based on the above method.
Fig. \ref{fig8} shows that with increasing $D$ up to the critical point
$D=D_c=1.114$ the plateau at $m=1/2$ is shrinking, while it is opening
again for $D>D_c$, in contrast to the classical system.
The plateau is due to the Haldane mechanism for $D<D_c$ while
the large-D mechanism for $D>D_c$.
This change of mechanisms should be a clear evidence of the quantum
nature in the plateau formation.

\section{Summary}

The magnetization process of the $(S,s)=(1,1/2)$ ferrimagnetic mixed
spin chain with the single-ion anisotropy $D$
was investigated using the numerical exact diagonalization
and some size-sailing analyses.
It revealed that the mechanism of the plateau at $m=1/2$
is changes from the Haldane to large-$D$ ones at the
Gaussian quantum critical point $D_c=1.114 \pm 0.001$.
It justifies that the plateau at $m=1/2$ originates from
the quantization of the magnetization,
although a similar plateau also appears in the classical system.

\section*{Acknowledgments}
It is a pleasure to thank Prof. S. Yamamoto for
helpful discussions.
This work was supported by the Japanese Ministry of Education,
Science, and Culture through Grant-in-Aid
for Scientific Research on Priority Areas (B) and  No. 13640371.
The numerical computation was done in part using the facility of the
Supercomputer Center, Institute for Solid State Physics, University of
Tokyo.

\widetext
\end{document}